\documentclass[twoside,twocolumn,9pt]{article}
\usepackage{extsizes}
\usepackage[super,sort&compress,comma]{natbib} 
\usepackage[version=3]{mhchem}
\usepackage[left=1.5cm, right=1.5cm, top=1.785cm, bottom=2.0cm]{geometry}
\usepackage{balance}
\usepackage{times,mathptmx}
\usepackage{sectsty}
\usepackage{graphicx} 
\usepackage{lastpage}
\usepackage[format=plain,justification=justified,singlelinecheck=false,font={stretch=1.125,small,sf},labelfont=bf,labelsep=space]{caption}
\usepackage{float}
\usepackage{fancyhdr}
\usepackage{fnpos}
\usepackage[english]{babel}
\usepackage{array}
\usepackage{droidsans}
\usepackage{charter}
\usepackage[T1]{fontenc}
\usepackage[usenames,dvipsnames]{xcolor}
\usepackage{setspace}
\usepackage[compact]{titlesec}

\usepackage[applemac]{inputenc}
\usepackage[T1]{fontenc}
\usepackage{placeins}

\usepackage{xr}
\usepackage{subcaption}
\usepackage{epstopdf}
\usepackage{accents}
\usepackage[
   cal = cm,
   bb = ams,
   frak = euler,
   scr = euler
]{mathalfa}

\definecolor{cream}{RGB}{222,217,201}

\newlength{\figrulesep} 
\title{Collective Dynamics of Self-propelled Semiflexible Filaments} 
\author{Ozer Duman$^{a}$, Rolf E. Isele-Holder$^{a}$, Jens Elgeti$^{a}$, and Gerhard Gompper$^{a}$} 
\begin{document}
\maketitle

\begin{abstract}
The collective behavior of active semiflexible filaments is studied with a model of tangentially
driven self-propelled worm-like chains. The combination of excluded-volume interactions and
self-propulsion leads to several distinct dynamic phases as a function of bending rigidity, activity,
and aspect ratio of individual filaments.
We consider first the case of intermediate filament density.
For high-aspect-ratio filaments, we identify a transition with increasing propulsion from a
state of free-swimming filaments to a state of spiraled filaments with nearly frozen
translational motion.
For lower aspect ratios, this gas-of-spirals phase is suppressed with growing density due to
filament collisions; instead, filaments form clusters similar to self-propelled rods, as
activity increases.
Finite bending rigidity strongly effects the dynamics and phase behavior.
Flexible filaments form small and transient clusters, while
stiffer filaments organize into giant clusters, similarly as self-propelled rods, but with a reentrant
phase behavior from giant to smaller clusters as activity becomes large enough to bend the
filaments.
For high filament densities, we identify a nearly frozen jamming state at low activities, a nematic
laning state at intermediate activities, and an active-turbulence state at high activities.
The latter state is characterized by a power-law decay of the energy spectrum as a function of 
wave number.
The resulting phase diagrams encapsulate tunable non-equilibrium steady states that can be used
in the organization of living matter.
\end{abstract}

\footnotetext{\textit{$^{a}$~Theoretical Soft Matter and Biophysics, Institute of Complex Systems and Institute for Advanced Simulations, Forchungszentrum J\"ulich GmbH, 52425, J\"ulich, Germany}}

\section{Introduction}

Living systems often self-organize into functional structures by consuming energy. Biopolymers and filamentous objects like actin filaments, microtubules and slender bacteria exhibit particularly interesting examples of self-organization,\cite{Karsenti_Self_2008, Kruse_Asters_2004, Loose_Spatial_2008, Couzin_Effective_2005, Riedel_A_2005, Nedelec_Self_1997, Prathyusha_Arxiv_2016, Elgeti_Physics_2015, Roland_Active_2017, Ravichandran_Enhanced_2017, kokot2017active} as their extended nature makes the collective dynamics inherently complex. Microtubules display loops when gliding on motility assays of kinesin-1 motors,\cite{Liu_Loop_2011} or on dynein-coated surfaces confined at an air-buffer interface.\cite{Kabir_Formation_2012} Likewise, large-scale vortices of microtubules on dynein carpets emerge due to inelastic collisions.\cite{Sumino_Large_2012, Ito_Formation_2014} Actin filaments self-organize into swirls at high densities when propelled by immobilized heavy meromyosin molecular motors.\cite{Schaller_Polar_2010} Besides these cytoskeletal filaments, other filamentous objects like slender bacteria \cite{Lin_Dynamics_2013} and synthetic particles such as vibrated granular rods are also found to self-organize into swirls.\cite{Blair_Vortices_2003, Kudrolli_Swarming_2008} Besides the fascinating physics of self-organization in active matter, understanding how these structures emerge from the underlying dynamics could shed light on their function. Cytoplasmic streaming of microtubules in the cortical arrays of plant cells provides an example, where the organization of microtubules into a swirl provides function in furnishing cell-wall growth.\cite{Chan_Cortical_2007} This type of self-organization can also play a useful role in micro- and nano-technology such as in nanofabrication and in drug delivery.\cite{Heuvel_Motor_2007}

Theoretical studies of active, flexible filaments have investigated different ways of invoking activity.\cite{Holder_Self_2015, Holder_Dynamics_2016, Jayaraman_Autonomous_2012, Jiang_Hydrodynamic_2014, Jiang_Motion_2014, Kaiser_2015, Eisenstecken_conformation_2016, Eisenstecken_Internal_2017, Kierfeld_2008} Activity, introduced as colored noise acting tangentially on a single filament, is shown to result in a net longitudinal drift of the filament.\cite{Liverpool_Anomalous_2003} In an ensemble of filaments with active colored noise acting over the normal direction of the bonds, the collective dynamics is observed to become superdiffusive with increasing levels of activity.\cite{Ghosh_Dynamics_2014} Such activity-caused enhanced diffusion can be rationalized with effective-temperature models.\cite{Loi_Effective_2011, Loi_Non_2011} 

Here we study the collective behavior of self-propelled semiflexible filaments with a focus on self-organization and dynamical pattern formation. We employ the self-propelled worm-like chain model, introduced recently to study the dynamics of a single semiflexible filament.\cite{Holder_Self_2015, Holder_Dynamics_2016} The self-propulsion is introduced as a constant magnitude force acting homogeneously in the tangential direction along the contour of the filament. 
The effect of self-propulsion has been shown to differ markedly for rigid and flexible filaments. 
It drives rigid filaments into a directed translational motion, where relaxation -- in particular rotational diffusion -- speeds up.
In contrast, when propulsion is stronger than bending rigidity,  filaments form spirals.\cite{Holder_Self_2015} 
For a finite-density suspension, we find that filaments cluster with increasing propulsion. Rigid filaments behave almost like rods, forming large clusters at intermediate propulsion. However, as propulsion increases, flexibility starts to play a role even for very stiff filaments and clusters break apart into smaller and highly motile clusters. At low rigidity, filaments coil up into a gas of isolated spirals if propulsion is sufficiently strong, as expected from single filament dynamics.
However, these spirals can be broken up by finite density effects if aspect ratio is to low.
For a high-density suspension, we find that filaments undergo a transition from a jammed state to a flowing state as a function of increasing activity. 
At intermediate levels of activity, filaments form nematic lanes, which break up into a active-turbulent regime upon further increase of activity.

\begin{figure*}[th!]
\centering
 \includegraphics[width=0.77\textwidth]{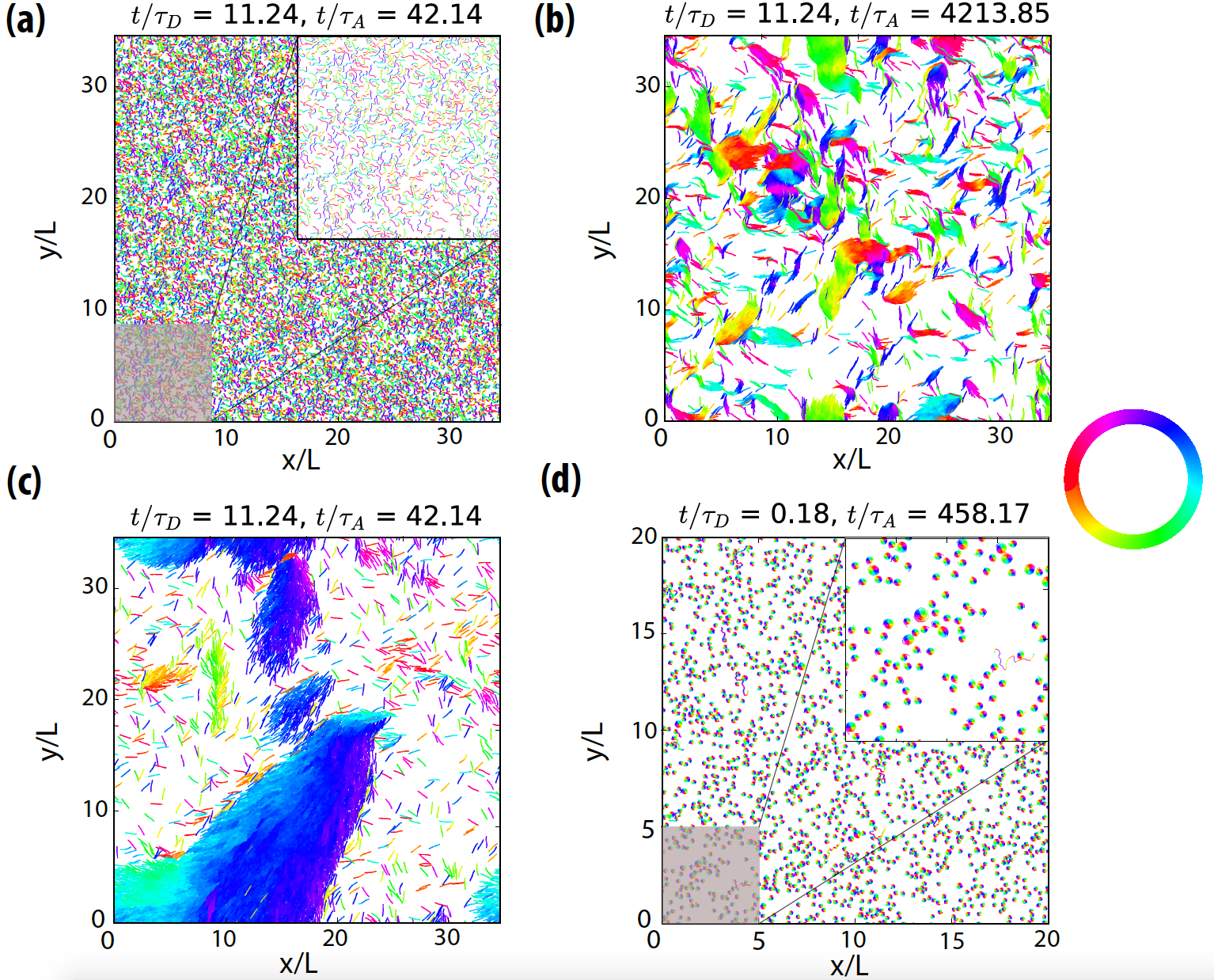}
 \captionsetup[subfigure]{justification=centering}
 \caption{Snapshots of the distinct phases: (a) melt phase characterized with a homogeneous distribution of filaments performing thermal motion ($\xi_{p}/L = 0.1$, $Pe = 15$, $a = 25$), (b) gas of clusters phase characterized with formation of small and transient clusters of filaments ($\xi_{p}/L = 16$, $Pe = 1500$, $a = 25$), (c) giant clusters phase characterized with formation of large and persistent clusters ($\xi_{p}/L = 16$, $Pe = 15$, $a = 25$), and (d) gas of spirals phase characterized with weakly interacting coiled filaments performing translational diffusion ($\xi_{p}/L = 0.1$, $Pe = 10000$, $a = 100$). Color wheel denotes orientation of each bond. Insets in (a) and (d) show zoomed areas shaded in gray.}
 \label{fgr:phase_picture}
\end{figure*}

\section{Model and methods}

Our study is based on the self-propelled worm-like chain model developed recently.\cite{Holder_Self_2015, Holder_Dynamics_2016} A single filament is represented by $N_{b} + 1$ beads held together by $N_{b}$ stiff bonds and bending potentials. The equation of motion is given by the Langevin equation
\begin{equation}
m\ddot{\mathbf{r}}_{i} = -\gamma \dot{\mathbf{r}}_i - \nabla_{i}U + \mathbf{F}_{k_{B}T}^{(i)} + \mathbf{F}_{p}^{(i)} ,
\end{equation}
where $\mathbf{r}_{i}$ are the coordinates of bead $i$ with the dots denoting derivatives with respect to time. $m$ denotes the mass, and $\gamma$ the friction coefficient of each bead. $U$ is the potential energy, $\mathbf{F}_{k_{B}T}^{(i)}$ is the thermal noise force acting on particle i, $ \mathbf{F}_{p}^{(i)}$ is the propulsion force, and $\gamma\dot{\mathbf{r}}_i$ is the drag force.

The configurational potential 
\begin{equation}
U = U_{\text{bond}} + U_{\text{bend}} + U_{\text{EV}}
\end{equation}
consists of bond, bending and excluded-volume contributions. The harmonic bond potential
\begin{equation}
U_{\text{bond}} = \frac{k_{S}}{2}\sum_{j}\sum_{i=1}^{N_{b}}(|\mathbf{r}_{i,i+1}^{(j)}| - r_{0})^{2} ,
\end{equation}
acts on the neighboring beads of each filament separately with $j$ and $i$ denoting the filament index and the bead index of a filament, respectively. The bond vector is defined as $\mathbf{r}_{i,i+1} = \mathbf{r}_{i+1}-\mathbf{r}_i$. $k_{s}$ is the spring constant for the bond potential and $r_{0}$ is the equilibrium bond length, which is the same for all the bonds. Semiflexibility is introduced via the bending potential,
\begin{equation}
U_{\text{bending}} = \frac{\kappa}{4r_0^2}\sum_{j}\sum_{i=1}^{N_{b}-1}\left(\mathbf{r}_{i,i+1}^{(j)} - \mathbf{r}_{i+1,i+2}^{(j)}\right)^2, 
\end{equation}
where $\kappa$ is the bending rigidity. The excluded-volume interaction acts between the beads of a single filament and between the beads of different filaments alike to render the filaments impenetrable. It is modeled by a Weeks-Chandler-Anderson potential, 
\begin{equation}
U_{\text{EV}} = \sum_{i}\sum_{j>i}u_{EV}(\mathbf{r}_{i,j}) , 
\end{equation}
\begin{equation}
u_{\text{EV}}(r) =
\begin{cases}
4\epsilon\lbrack \left(\frac{\sigma}{r}\right)^{12} - \left(\frac{\sigma}{r}\right)^6\rbrack+ \epsilon, & r < 2^{1/6}\sigma \\
0, & r \geq 2^{1/6}\sigma
\end{cases}
\end{equation}
where $\mathbf{r}_{i,j} = \mathbf{r}_{i} - \mathbf{r}_{j}$ is the vector between the positions of the beads i and j (which may belong to the same or different filaments). $\epsilon$ and $\sigma$ are the characteristic volume-exclusion energy and effective filament diameter, respectively.

Self-propulsion is introduced as a constant magnitude force acting along each bond of a filament tangentially, i.e., $\mathbf{F}_{p}^{(i)} = f_{p}\mathbf{r}_{i,i+1}$. This force is distributed equally among both adjacent beads constituting the bond. The thermal force $\mathbf{F}^{(i)}_{k_{B}T}$ is modeled as white noise with zero mean and variance $2k_{B}T\gamma /\Delta t$, to fulfill the fluctuation-dissipation theorem. The parameter choice is done such that (i) $k_{s}$ is sufficiently large to render the bond length essentially constant at $r_{0}$, and (ii) the local filament curvature low such that the bead discretization does not violate the worm-like chain description.\cite{Holder_Self_2015} When the aforementioned conditions are met, the dynamics of a single filament is described by two dimensionless numbers,
\begin{equation}
\xi_{P}/L = \frac{\kappa}{k_{B}TL} ,
\end{equation}
\begin{equation}
Pe = \frac{f_{p}L^2}{k_{B}T} ,
\end{equation}
where $L = N_{b}r_{0}$ is the filament length (with $N_{b}$ denoting the number of bonds of a filament) and $\xi_{p}$ is the persistence length of the chain. The Peclet number $Pe$ is the ratio between advective and diffusive transport, thereby providing a measure for the strength of self-propulsion. Two other parameters related to the size of the filament can also play a role.\cite{Holder_Self_2015, Abkenar_Collective_2013} The aspect ratio $a = L/\sigma$ is the ratio of the length of the filament to the effective diameter of a bead. Furthermore, the smoothness of the filament, measured by the degree of overlap of neighbouring beads $\sigma/r_0$, determines an effective friction between different segments in close contact.

\begin{figure}[h]
\centering
 \includegraphics[width=0.9\columnwidth]{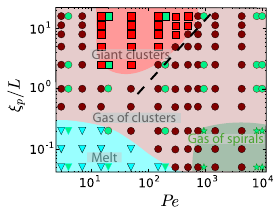}
 \caption{Phase diagram as a function of $\xi_p/L$ and $Pe$. Symbols encode the distinct phases. Cyan triangles display the melt phase, magenta circles are the gas of clusters phase and the red squares are the giant clusters phase. The background colors are guides to the eye. Phases are identified with the average cluster size. Points with an average cluster size smaller than 10 filaments are marked as the melt phase, while those larger than 200 filaments are marked as the giant clusters, with everything in between falling into the category of gas of clusters. Green filled symbols depict the simulations with high aspect ratio filaments (at $a=100$, while every other points denote simulations with $a=25$). The only different phase for high aspect ratio filaments is the gas-of-spirals phase, which is depicted with stars with every other symbol remaining the same as before. The black dashed line indicates that the phase boundary between the giant clusters and gas of clusters phases is well described by the linear relation $\xi_p/L \sim Pe$, \emph{i.e.} flexure number $\mathfrak{F} \approx 100$.}
 \label{fgr:phase_diagram}
\end{figure}

The contour velocity of a filament is given by $v_{c} = f_{p}/\gamma_{l}$, and the translational diffusion coefficient $D_{t} = k_{B}T/\gamma_{l}L$. The friction coefficient per unit length is given with $\gamma_{l} = \gamma(N_{b}+1)/L$. The flexure number
\begin{equation}
\mathfrak{F} = PeL/\xi_{p} = \frac{f_{p}L^{3}}{\kappa}
\end{equation}
is the ratio between activity and bending rigidity.

The results are presented in dimensionless units, with length measured in units of the filament length $L$, energies in units of the thermal energy $k_{B}T$, 
and time in units of the self-diffusion time $\tau_D$ for the filament to diffuse its own body length (diffusive time) or the self-advection time $\tau_A$ for the filament to propel itself along its own body length (advective time), with
\begin{equation}
\tau_{D} = L^{3}\gamma_{l}/4k_{B}T ,
\end{equation}
\begin{equation}
\tau_{A} = L\gamma_{l}/f_{p}.
\end{equation}

Equations of motion are integrated with the Verlet algorithm using LAMMPS.\cite{Plimpton_1995} The usage of Verlet integration instead of Euler integration allows for a larger timestep and therefore provides faster computations. $m$ and $\gamma$ are chosen such that the dynamics is close to overdamped dynamics ($m/\gamma = 1.2 \times 10^{-5}\tau_D$). To ensure that inertial effects are negligible for the presented results, we rerun some of the simulations at different values of friction coefficient and observe the same behavior.

Unless otherwise stated, we use $k_{s} = 5000k_{B}T/\sigma_{0}^{2}$, $r_{0} = \sigma/2 = L/N_{b}$, and $\epsilon = k_{B}T$. We consider a square box of lengths $L_{x} = L_{y} \approx 866\sigma$ with periodic boundary conditions. We define the packing fraction as $\phi = N_{b}N_{f}\sigma r_{0}/L_{x}L_{y}$. $N_{b}$ denotes the number of bonds making up a single filament, $N_{f}$ the number of filaments. We use $N_f = 6000$ for low aspect ratio filaments ($a = 25$) and $N_f = 2000$ for high aspect ratio filaments ($a = 100$) with $\phi = 0.2$ and $\phi = 0.8$. The parameter space is explored via changing $N_{b}$, $f_{p}$ and $\kappa$ to vary the aspect ratio $a$, Peclet number $Pe$ and the thermal persistence length $\xi_{P}/L$, respectively.  

\begin{figure*}[t]
\centering
 \includegraphics[width=0.8\textwidth]{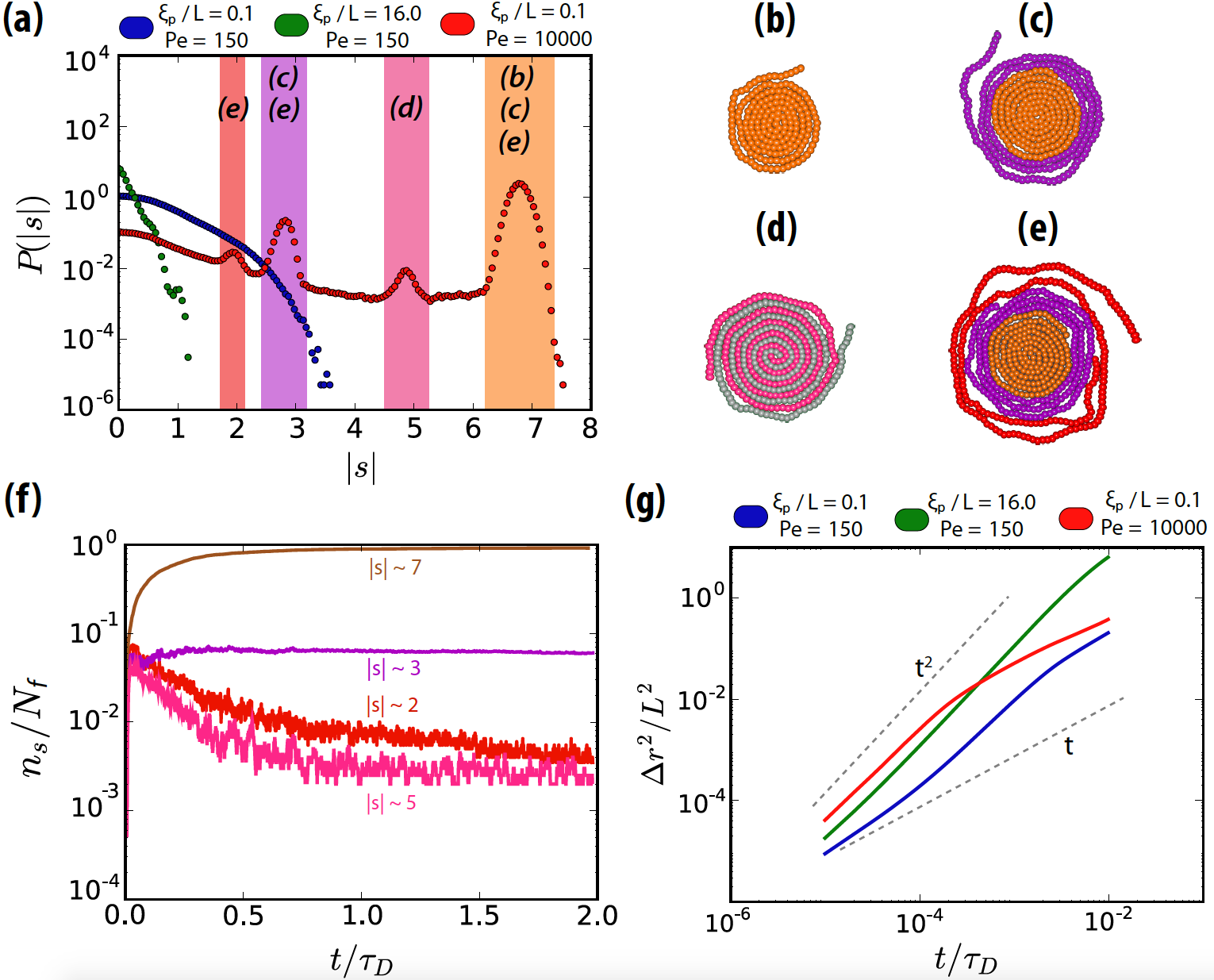}
 \captionsetup[subfigure]{justification=centering}
 \caption{(a) Probability distribution of the absolute value of the
   spiral number. (b-e) Distinct spiral configurations observed in the
   system: (b) A single coiled filament with $|s|\approx 7$, (c) a
   filament wrapped around a coiled filament, with $|s|\approx 3$, (d)
   two intertwined filaments each with $|s|\approx 5$, and (e) two
   filaments wrapped around a coiled filament, with $|s|\approx 3$ and
   $|s|\approx 2$, respectively. The color codes of filaments are
   consistent with the colors of the bars around the peaks in (a). (f)
   Time evolution of number of filaments with approximately the given
   spiral numbers $n_s$. The intervals for each $|s|$ value are
   colored with a bar around each peak of (a). The parameters are
   $\xi_p/L=0.1, Pe=10000$. (g) Mean squared displacement of the
   center of mass of filaments. Aspect ratio is $a= 100$ (see also
   Movie M5 and M6 in the ESI). }
 \label{fgr:spiral_set}
\end{figure*}

\section{Results}

The collective dynamics of self-propelled semiflexible filaments is dictated by the dynamics of intra- and inter-filament collisions, which are in turn dependent on the properties of a single filament. The dynamical behavior of a single filament is determined by its thermal persistence length $\xi_p/L$, Peclet number $Pe$ and aspect ratio $a$.\cite{Holder_Self_2015} At high $\xi_p/L$, the active forces drive the filament along its contour leading to directed translational motion. However, increasing activity also increases the flexure number, \emph{i.e.} the activity becomes able to deform the filaments. 
At high aspect ratio, filaments spontaneously form spirals in this
regime of high $Pe$ combined with low
$\xi_p/L$.\text{\cite{Holder_Self_2015}} We distinguish four distinct
phases in the collective dynamics at finite filament concentration
(see Fig.~\ref{fgr:phase_picture}). The region of stability of each of
these non-equilibrium phases is depicted in the phase diagram in
Fig.~\ref{fgr:phase_diagram}. At low $\xi_p/L$ and $Pe$, the ensemble
behaves like a passive homogeneous melt
(Fig. \ref{fgr:phase_picture}-a). Increasing $Pe$ leads to the
formation of clusters. Clusters of low $\xi_p/L$ filaments are
transient and small (Fig. \ref{fgr:phase_picture}-b), while clusters
of high $\xi_p/L$ filaments are large, persistent and rotating
(Fig. \ref{fgr:phase_picture}-c and Movie M2 in the ESI). However, we
find that cluster sizes do not monotonically grow in $Pe$, but instead
peak at moderate propulsion strengths (see Movie M3 in the ESI). 

\subsection{Spiral Formation}
Isolated filaments wind up in spirals at strong propulsion and low rigidities. 
Spiral formation is facilitated by self-interactions. 
It is initiated by the head of the filament colliding with a subsequent part of its body. 
As such, the probability of spiral formation depends on activity and persistence length. 
However, once the spirals are formed, their stability is altered by aspect ratio.
Filaments wind up over themselves more with increasing aspect ratio. 
Therefore, spirals of higher aspect ratio filaments are harder to break apart. 
This role of aspect ratio becomes crucial for the collective dynamics. 

We begin our analysis with the dynamics of high-aspect-ratio filaments. To elucidate spiral formation and spiral structure, we calculate the spiral number
\begin{equation}
s = \sum_{j=1}^{N_f}\frac{\theta_j(L) - \theta_j(0)}{2\pi N_f} , 
\end{equation}
where $\theta_j(l)$ is the bond orientation at position $l$ along the contour of the filament. It is a quantitative measure of the number of times the filament has wrapped around its head bead. 

For filaments with low propulsion and low rigidity (at $\xi_p/L = 0.1$, $Pe=150$), the probability distribution of the absolute value of the spiral number resembles a Gaussian distribution cut in half (see Fig.~\ref{fgr:spiral_set}-a). 
With increasing rigidity (at $\xi_p/L = 16$, $Pe = 150$), the distribution becomes exponential and develops a small peak around $|s|\approx 1$. Stiff filaments are bent into circles (corresponding to $|s|\approx 1$) due to the stress generated by self-propulsion. However, such circular loops are transient in dynamics, so that the peak is small. 

Strongly propelled filaments with low rigidity (at $\xi_p/L=0.1$, $Pe=10000$) self-organize into spirals of different configurations. This results in multiple peaks in the probability distribution of $|s|$. Each peak denotes a different type of spiral configuration that are highlighted in Fig.~\ref{fgr:spiral_set}-b to e. The most pronounced peak, coinciding with the most common type of spiral in the ensemble, is around $|s|\approx 7$, corresponding to a single coiled filament. When the leading tip of a filament hits a subsequent part of its own body, the volume exclusion force bends the leading tip, thereby causing it to wind onto itself. After the winding process, the filament stays locked in the coiled form for an extended duration. The number of spirals consisting of a single filament increases up to a value close to unity in time (see Fig.~\ref{fgr:spiral_set}-f), rendering the dynamics of the ensemble analogous to a gas of spirals. This is consistent with the dynamics of a single filament.\text{\cite{Holder_Self_2015}}

As a direct result of collective motion, there are multiple spiral configurations corresponding to the different peaks in the distribution. The second important peak occurs at $|s|\approx 3$. The configuration it represents is involved in two types of spirals (in double-wrapped and triple-wrapped filament configurations in Figs.~\ref{fgr:spiral_set}-c and e). The double-wrapped filament configuration consists of a single coiled filament with another filament wrapped around it. Another filament can wind itself around these two filaments to form the configuration with three filaments. The spiral number of the third wrapping filament coincides with the first peak of the distribution. It is relatively easier to break up this state. As a result, number of filaments with $|s|\approx 2$ decreases with time. 

Finally, the smallest peak (at $|s| \approx 5$) represents a configuration with two intertwined filaments (Fig.~\ref{fgr:spiral_set}-d). Two filaments interlock when one of them hits the other before winding itself over it. This type of formation has the lowest probability to occur. It is also easy for the filaments to break out of this interlock, as a free filament can widen the tail of one of the interlocking filaments, thereby facilitating the break-up. As a result, the observed peak around $|s|\approx 5$ is small.

In terms of dynamical behavior, spirals rotate in accordance with their wrapping direction, with the center of mass performing thermal motion. The mean square displacement (MSD) of the center of mass of long filaments in the spiralling phase is marked by an initial ballistic behavior ($\sim t^2$) that makes a transition into a diffusive behavior ($\sim t$) at late lag times (see Fig.~\ref{fgr:spiral_set}-g). The initial ballistic regime is a result of filaments performing directed motion at early lag times. Therefore, the transition time from ballistic to diffusive motion gives a time scale for spiralling on average. The movement of the filaments is significantly hindered in the spiralling regime, despite the strong propulsion. 

\begin{figure}[h]
\centering
 \includegraphics[width=\columnwidth]{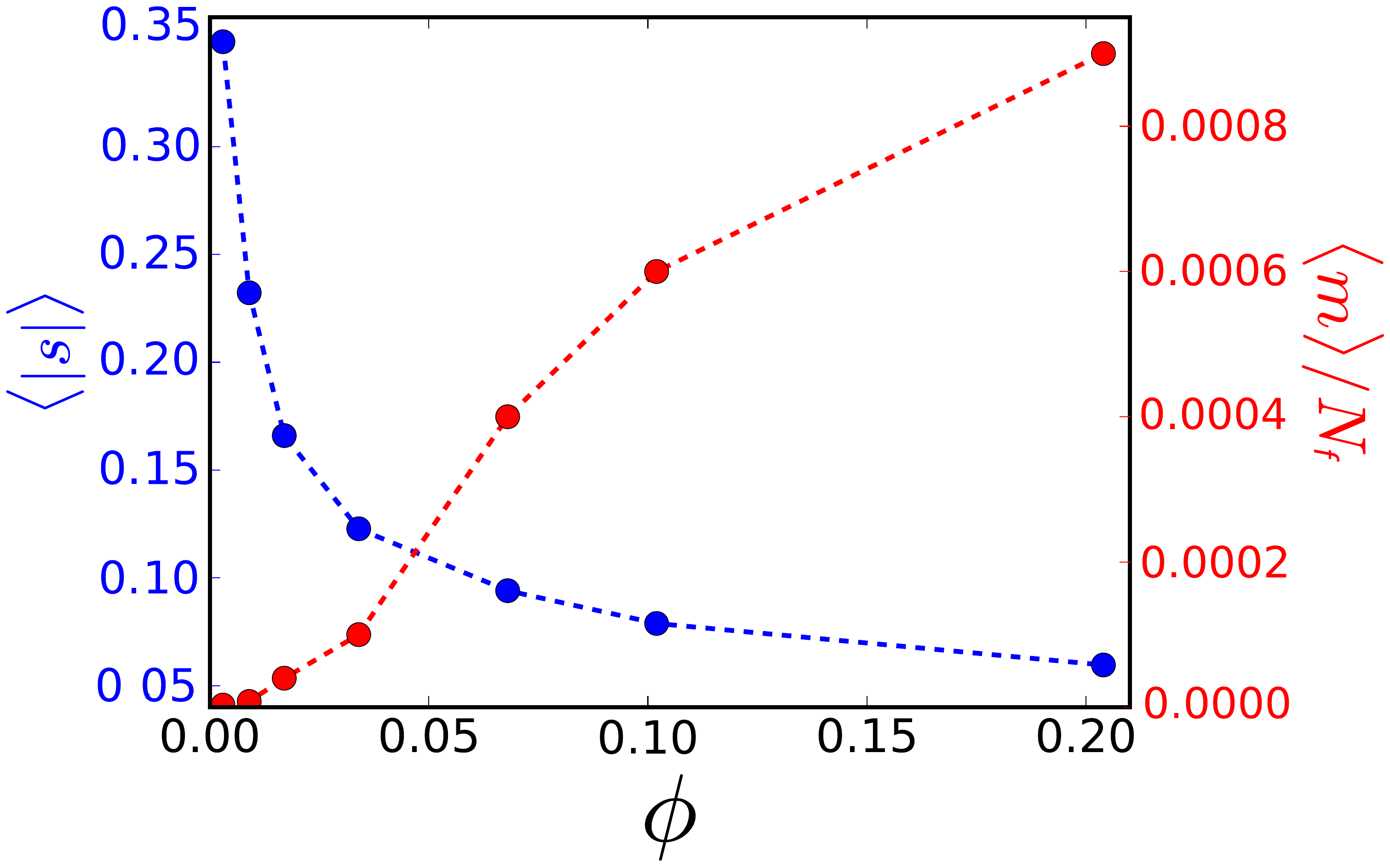}
 \caption{The average of the absolute value of spiral number (blue-shaded axis) and the average cluster size (red-shaded axis) as a function of density. Density is varied by changing the number of filaments ($N_f = 100, 250, 500, 1000, 2000, 3000, 6000$ for $\phi = 0.003, 0.009, 0.017, 0.034, 0.068, 0.102, 0.204$, respectively). The other parameters are fixed at $\xi_p/L = 0.1$, $Pe = 4300$, and $a = 25$.}
 \label{fgr:sdensity}
\end{figure}

For lower aspect ratios, collective effects radically change the picture.
Spirals of shorter filaments ($a=25$) are winded a smaller number of times, which renders them less stable. 
At low concentration, these filaments are in the spiraled state most of the time. 
However, as density increases, collisions of uncoiled filaments with
spirals break the spirals (see Movie M5 in the ESI). Thus the average spiral number decreases with filament density (see Fig. \ref{fgr:sdensity}) until spirals are almost absent at $\phi=0.2$.
Instead, the filaments form increasingly large motile clusters. 

\subsection{Cluster Formation and Disintegration}

We consider two filaments as part of the same cluster if they have closely spaced bonds that are pointing in the same direction. More specifically, if $30\%$ of the body of two filaments are within a distance range of $2\sigma$ of each other, and point in the same direction $\pm \pi/6$, then we consider the filaments to be part of the same cluster. The average cluster size is then defined as $\langle m\rangle = \sum_{m} m\langle P(m,t)\rangle$ with the average taken over time.

\begin{figure}[h]
\centering
 \includegraphics[width=0.9\columnwidth]{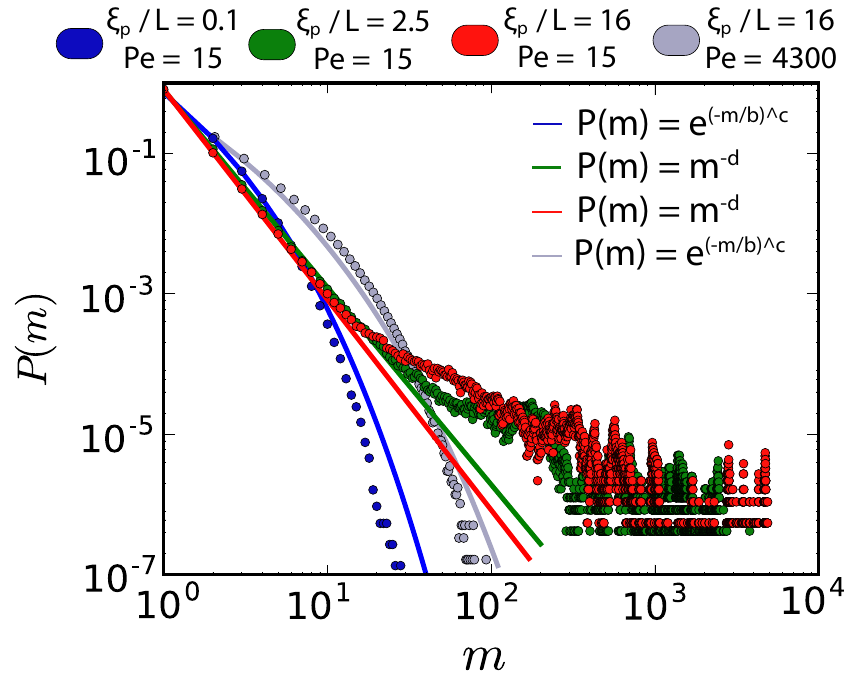}
 \caption{Cluster mass distribution as a function of $\xi_p/L$ at $Pe = 15$ and $Pe = 4300$ as depicted in the outer legend. Aspect ratio is fixed at $a = 25$. Solid lines are fits as given in the inner legend. The values of fit parameters are given in the text.}
 \label{fgr:cluster_mass_dist}
\end{figure}

To quantify the transition into clustering, we calculate the cluster mass distribution, 
\begin{equation}
P(m,t) = \frac{mn_{m}(t)}{N_{f}} ,
\end{equation}
where $n_{m}(t)$ is the number of clusters of mass $m$ present at time $t$ and $N_{f}$ is the total number of filaments in the system. The distribution fits a stretched exponential of the form $P(m) = \exp(-m/b)^c$ with $b = 0.02$ to $0.04$ and $c = 0.2$ to $0.4$ for filaments with low rigidity and low propulsion (see Fig.~\ref{fgr:cluster_mass_dist}). The stretched exponential fit points to a dynamical clustering picture where clusters of different mass acquire and eject filaments at different characteristic rates.\cite{Glotzer_Dynamical_1998}. With an increasing rigidity, the cluster mass distribution undergoes a transition to a power law distribution of the form $P(m) = m^{-d}$ with $d\approx 3$. Additionally, the distribution has a nonmonotonically decreasing tail with a shoulder at large mass values, indicating an increase in the probability of finding a filament in a larger cluster. This corresponds to the giant cluster regime in Fig. \ref{fgr:phase_diagram}. However, as the activity is increased to $Pe = 4300$, the distribution turns back to a stretched exponential with $b\approx0.07$ and $c\approx0.3$. Remarkably, formation of large clusters disappear with increasing activity. 

\begin{figure}[h]
\centering
 \includegraphics[width=0.9\columnwidth]{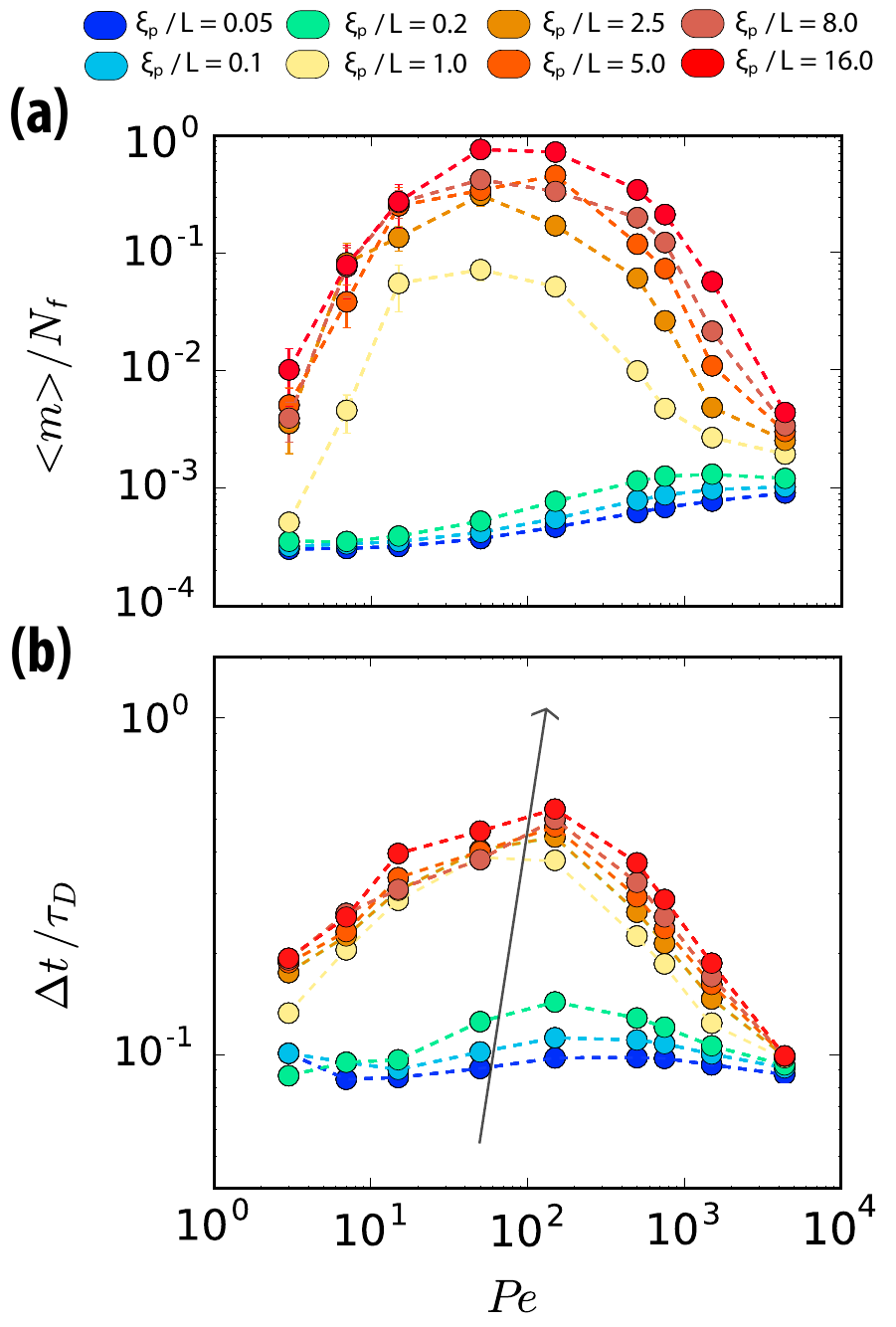}
 \caption{(a) Average cluster mass (in terms of bead mass), and (b) average lifetime of clusters as a function of $Pe$. Different $\xi_p/L$ values are denoted in the legend. The arrow direction indicates increasing $\xi_p/L$. Aspect ratio is $a = 25$.}
 \label{fgr:cluster_averages}
\end{figure}

To elucidate the properties of the clusters further, we calculate the average cluster size and the average lifetime of clusters. For the average lifetime calculation, we label the clusters in time by comparing the identity of filaments in two successive time frames. The clusters with maximum overlapping filaments are labelled as the same cluster with a unique identity. A cluster is deemed to lose its identity when its size drops down to less than 10 filaments or when it merges with a larger cluster. The tendency of clustering decreases with decreasing rigidity. However, for increasing activity, it reaches a peak at intermediate values, after which it starts to decrease. Therefore, both average cluster size and average cluster lifetime have distinct peaks at intermediate $Pe$ for stiff filaments (see Fig.~\ref{fgr:cluster_averages}). The maximum at intermediate activity level corresponds to the pocket of giant clusters in the phase diagram (Fig.~\ref{fgr:phase_diagram}).

The disappearance of giant clusters at high $Pe$ is a flexibility effect, as stiff rods do not show this reentrant behavior. Two factors contribute to cluster disappearance. Dynamics of self-propelled filaments are sped up, leading to faster reorientation, and active forces are deforming the filaments themselves.

\begin{figure}[h]
\centering
 \includegraphics[width=0.9\columnwidth]{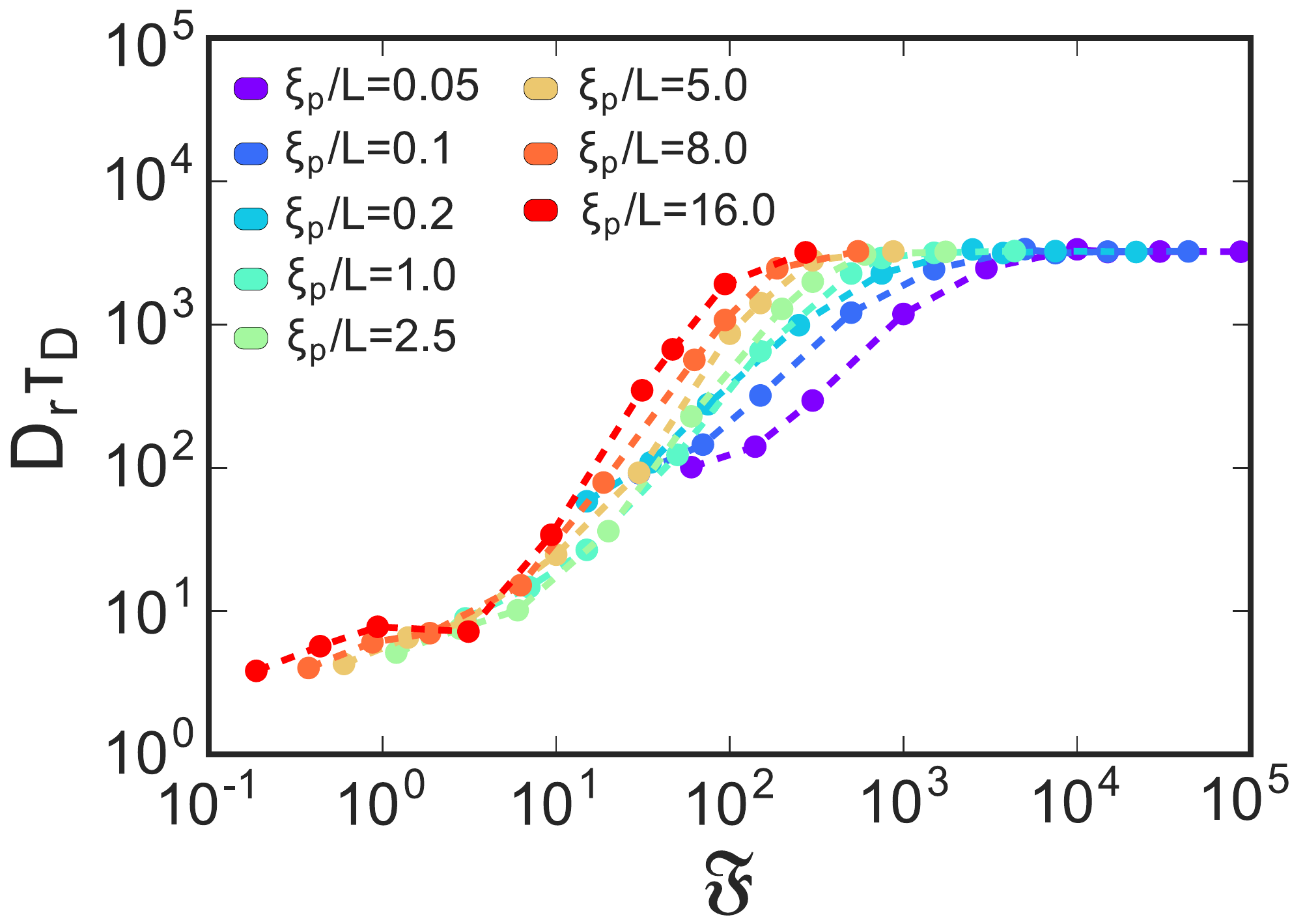}
 \caption{Rotational diffusion coefficients as extracted from the MSR
   of the end-to-end vector of filaments as a function of flexure number $\mathfrak{F}$ for different
 $\xi_p/L$ as indicated in the legend. Aspect ratio is $a = 25$.}
 \label{fgr:dr}
\end{figure}

 The rotational diffusion coefficients can be extracted from the
 average orientation of the end-to-end vector $\theta$ of
 filaments. Fitting the mean square rotation (MSR) of the end-to-end
 vector with $\Delta \theta^2 = 2D_{r}t$ 
yields the effective rotational diffusion coefficient
 $D_r$. As activity increases, rotational diffusion increases (see Fig.~\ref{fgr:dr}).
This increased rotational diffusion does not arise from the railway
motion of isolated filaments, which sets in at higher flexure numbers ($\mathfrak{F} > 100$).\cite{Holder_Self_2015} Instead, the rotational diffusion displays also collectivity effects.
 Filaments have a much shorter mean free path at high $Pe$, which causes them to reorient very frequently by bending upon collisions. As a result, the rotational diffusion of filaments gets a significant activity-enhanced contribution, which destroys large clusters. 

\begin{figure}[h]
\centering
 \includegraphics[width=0.9\columnwidth]{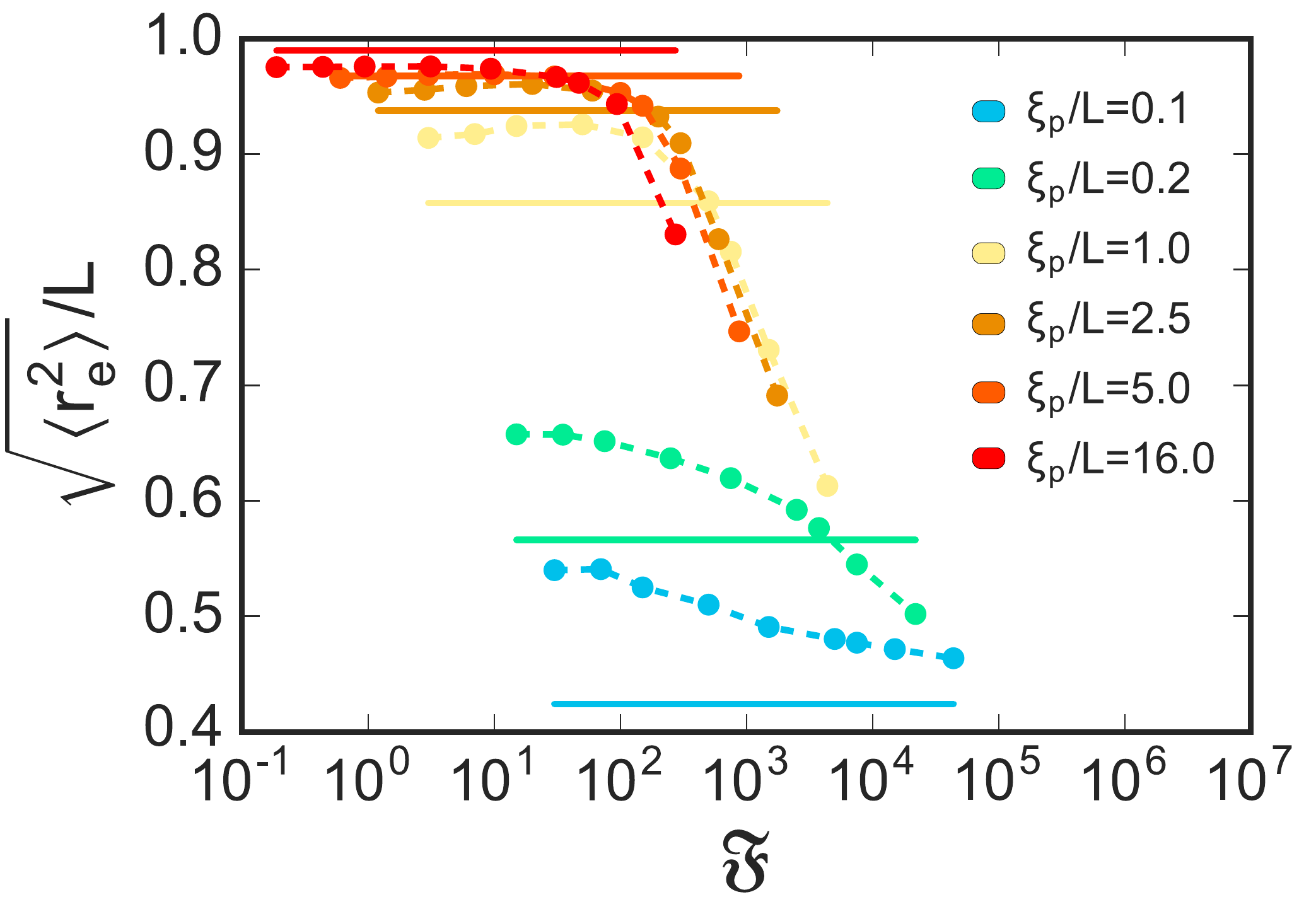}
 \caption{Average end-to-end distance $\sqrt{\langle r_e^2\rangle}$ of
   filaments with respect to flexure number $\mathfrak{F}$ for different
 $\xi_p/L$ as indicated in the legend. Solid lines show the theoretical values from the Kratky-Porod model. Aspect ratio is $a = 25$.}
 \label{fgr:e2e}
\end{figure}

We calculate the average end-to-end distance $\sqrt{\langle r_e^2\rangle}$ to quantify the structural properties of the filaments. The Kratky-Porod worm-like chain model, valid for passive filaments without volume exclusion, predicts\cite{Kratky_Diffuse_1949}
\begin{equation}
\frac{\langle r_e^2\rangle}{L^2} = 2\frac{\xi_p}{L} - 2\bigg(\frac{\xi_p}{L}\bigg)^2\big(1-\mathrm{e}^{-L/\xi_p}\big)
\label{eqn:kratky_porod}
\end{equation}
The end-to-end distances generally deviate from the Kratky-Porod
prediction due to the excluded-volume interactions, especially for low
rigidities. However, for a
single filament, the end-to-end distance is found to agree well with
the Kratky-Porod prediction at high rigidities, significant
deviations occur at low rigidities due to spiral
formation.\cite{Holder_Self_2015} 
For the collective dynamics of stiff filaments ($\xi_p>L$), we find that the end-to-end distance $\sqrt{r_e^2}$ is nearly constant up to a
critical  flexure number $\mathfrak{F} \approx 100$, then starts to decrease approximately logarithmically
(Fig. \ref{fgr:e2e}). 
This behavior is consistent with the dependence of the phase boundary of the giant-cluster phase at higher $Pe$ in Fig. \ref{fgr:phase_diagram}, which shows a similar linear dependence. Both of these observations indicate that the reentrant behavior in Fig. \ref{fgr:phase_diagram} is due to a balance of propulsion and curvature forces. The giant clusters dissolve when propulsion forces ($\sim Pe$) become strong enough to significantly deform the filaments with bending-induced restoring forces ($\sim \xi_p/L$), \emph{i.e.} flexure number $\mathfrak{F} \approx 100$.

\subsection{Dynamics of Giant Clusters}

\begin{figure}[h]
\centering
 \includegraphics[width=0.9\columnwidth]{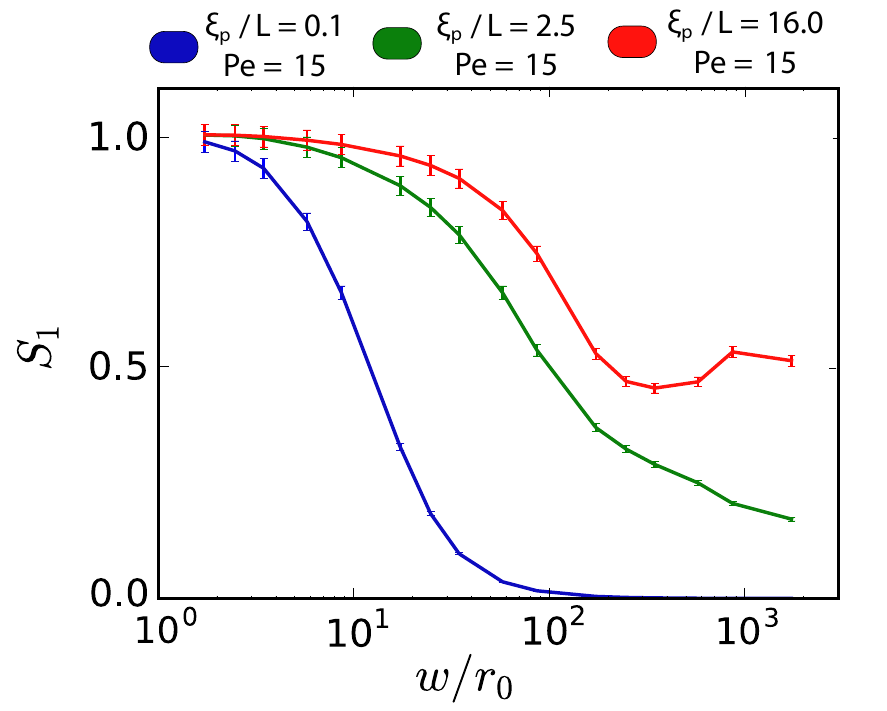}
 \caption{Polar order parameter inside square boxes with respect to different box sizes. Colors correspond to different $\xi_p/L$ values as given in the legend. $Pe=15$ and $a=25$ in all three curves.}
 \label{fgr:polar_order}
\end{figure}

Giant clusters consist of polar ordered filaments. They occur in a part of the phase space wherein filaments are stiff ($\xi_p/L \ge 2$) and the propulsion force is at an intermediate value ($10^1 \le Pe \le 10^3$). We calculate the polar order parameter of the bond vectors,
\begin{equation}
S_{1}(w) = \langle|\langle e^{i\theta_{j}(t)}\rangle|\rangle ,
\end{equation}
where j denotes the bond identities within boxes of area $w^{2}$, and the averages are taken over bonds and time. In this way, we calculate the polar order locally inside square boxes and study it as a function of box size (see Fig.~\ref{fgr:polar_order}). As filaments become stiffer at intermediate activity, clusters of polar ordered filaments span larger and larger areas. The increase in polar ordering can be attributed to the alignment of stiff filaments upon collisions with the same mechanism as collision-induced alignment of self-propelled rigid rods.\cite{Weitz_Self_2015}

\begin{figure}[h]
\centering
 \includegraphics[width=0.9\columnwidth]{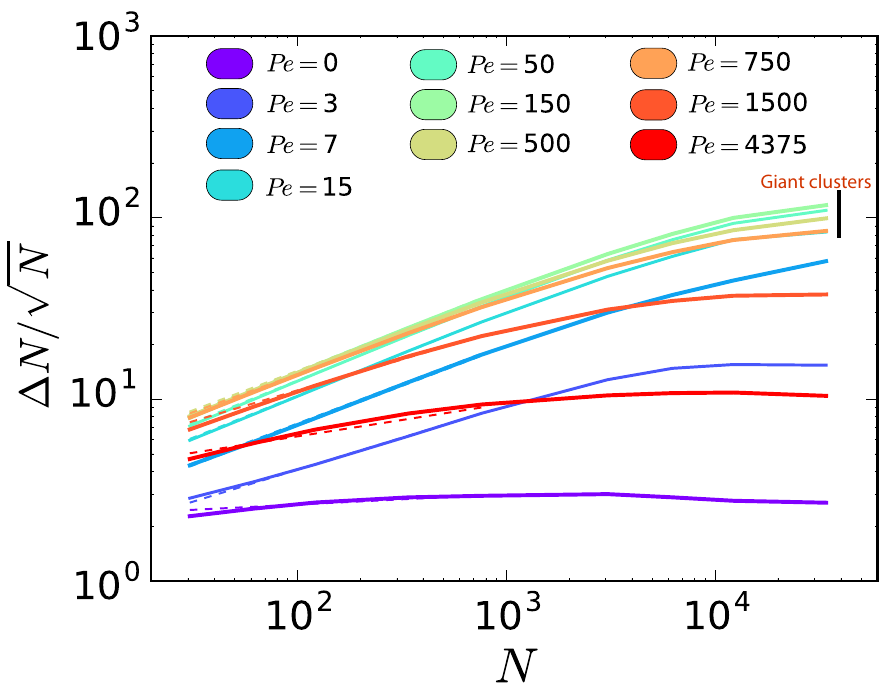}
 \caption{Number fluctuations of beads in square boxes with respect to the average number of beads inside the box. The persistence length is fixed at $\xi_p/L = 16$ and aspect ratio at $a = 25$, while changing $Pe$ is given in the legend. Power law fits are shown in the dashed lines. Giant clusters regime is marked with black line.}
 \label{fgr:number_fluc}
\end{figure}

Highly polar-ordered phases of self-propelled rods are observed to
exhibit giant number fluctuations.\cite{Ginelli_Large_2010,
  Peruani_Collective_2012, Peruani_Active_2016, Narayan_Long_2007}
Such fluctuations are given by the variance of the number of particles
($\Delta n^2 = \langle n^2\rangle - \langle n\rangle^2$) within square
boxes of varying sizes. For thermal motion accompanied with a
homogeneous density distribution, the number fluctuations scale with
$\Delta n^2 \approx n$, where $n$ is the average number of
particles. We observe an increase in the power $\alpha$ in $\Delta n
\approx n^{\alpha}$ with increasing $Pe$ when the rigidity is high
($\xi_p/L = 16$, see Fig.~\ref{fgr:number_fluc}). It increases from $\alpha = 0.5$ at $Pe = 0$ up to $\alpha = 0.9$ at $Pe = 150$, which corresponds to the mid-part of the giant clusters regime. When $Pe$ increases further, $\alpha$ decreases again all the way down to $\alpha = 0.6$ at $Pe = 4375$ (for comparison, a power of $0.8$ is observed for rods in the giant-cluster regime).\cite{Ginelli_Large_2010, Peruani_Collective_2012} Therefore, the giant-cluster regime is marked with giant number fluctuations due to the density inhomogeneities caused by clustering. 

In forming giant clusters, active semiflexible filament ensembles
behave similarly as ensembles of self-propelled rigid rods. However,
even at high rigidities, flexibility plays a crucial role. One
interesting manifestation of flexibility is in the collisions of giant
clusters. For a dense system of active rigid rods, head-on collisions
of large clusters are observed to lead to an accumulation of stress
resulting in the formation of a jammed circular
aggregate.\cite{Weitz_Self_2015} In an ensemble of active semiflexible
filaments, on the other hand, filaments, either individually or in
clusters, can open up channels inside the cluster they are hitting
head on. They can bend their way through the antagonistic cluster
until they escape (See Movie M2 in the ESI).

Another interesting aspect of giant clusters is the ease of
orientation-angle transmission between the filaments of a
cluster. When the leading portion of filaments in the front part of a
cluster change orientation, either due to a collision or thermal
diffusion, the other filaments follow suit in reorienting themselves
in the new direction and thereby they turn the entire cluster (see Movie M4 in the ESI). As a
result, giant clusters are rotating frequently and their shape is
often curved and meandering 
(see Movie M2 in the ESI). 
\subsection{Dynamics at High Densities}

\begin{figure*}[t]
\centering
 \includegraphics[width=0.95\textwidth]{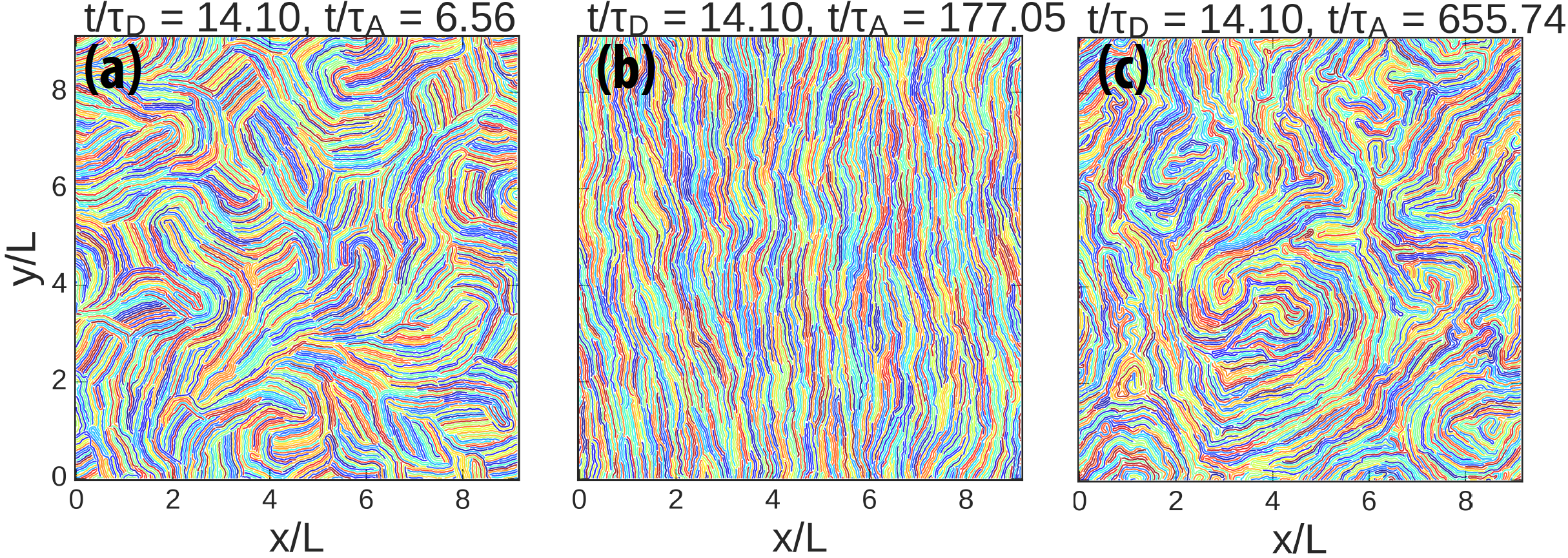}
 \captionsetup[subfigure]{justification=centering}
 \caption{Snapshots of high density phases at $\phi = 0.8$, $a = 30$, $\xi_p/L = 0.2$ for various $Pe$: (a) Jamming ($Pe = 0.9$), (b) laning ($Pe = 24.3$), and (c) active turbulence phases ($Pe = 90$). The color code is chosen as random to discern individual filaments.}
 \label{fgr:high_dens_phase_picture}
\end{figure*}

Filament motion becomes increasingly hindered at high densities.
At a high enough density ($\phi = 0.8$), even flexible filaments get
jammed due to the high energy barriers associated with structural
rearrangements (Fig. \ref{fgr:high_dens_phase_picture}-a and Movie M7 in the ESI). 
In the jamming regime (low activity) stacks of filaments displace by
small amounts with respect to other stacks towards energetically more
favourable configurations. 
With increasing activity filaments overcome the energy barriers and
rearrange.
In the initial stage of the simulation, the dynamics are now dominated by
the creation and annihilation of half-integer topological defects due
to the nematic symmetry of filaments, similar to the defect behavior
in active-nematic fluids.\cite{DeCamp_Nat_2015, Hemingway_Active_2015} 
However,  for intermediate levels of activity,
 we find that this defect-rich state is only transitory and
relaxes into a nematic laning regime in the steady state
(Fig. \ref{fgr:high_dens_phase_picture}-b and Movie M8 in the ESI). 
At high activity, active turbulence constitutes the steady-state
dynamics as the propulsion forces decrease the effective persistence
length, and thereby the nematic correlations in motion. Thus, fully nematic lanes break up into smaller nematic-aligned bands of moving filaments. The dynamics is then dominated by the collisions between these nematic bands, which results in an active turbulence phase with continuous creation and annihilation of defects (Fig. \ref{fgr:high_dens_phase_picture}-c and Movie M9 in the ESI).   

\begin{figure}[h]
\centering
 \includegraphics[width=0.95\columnwidth]{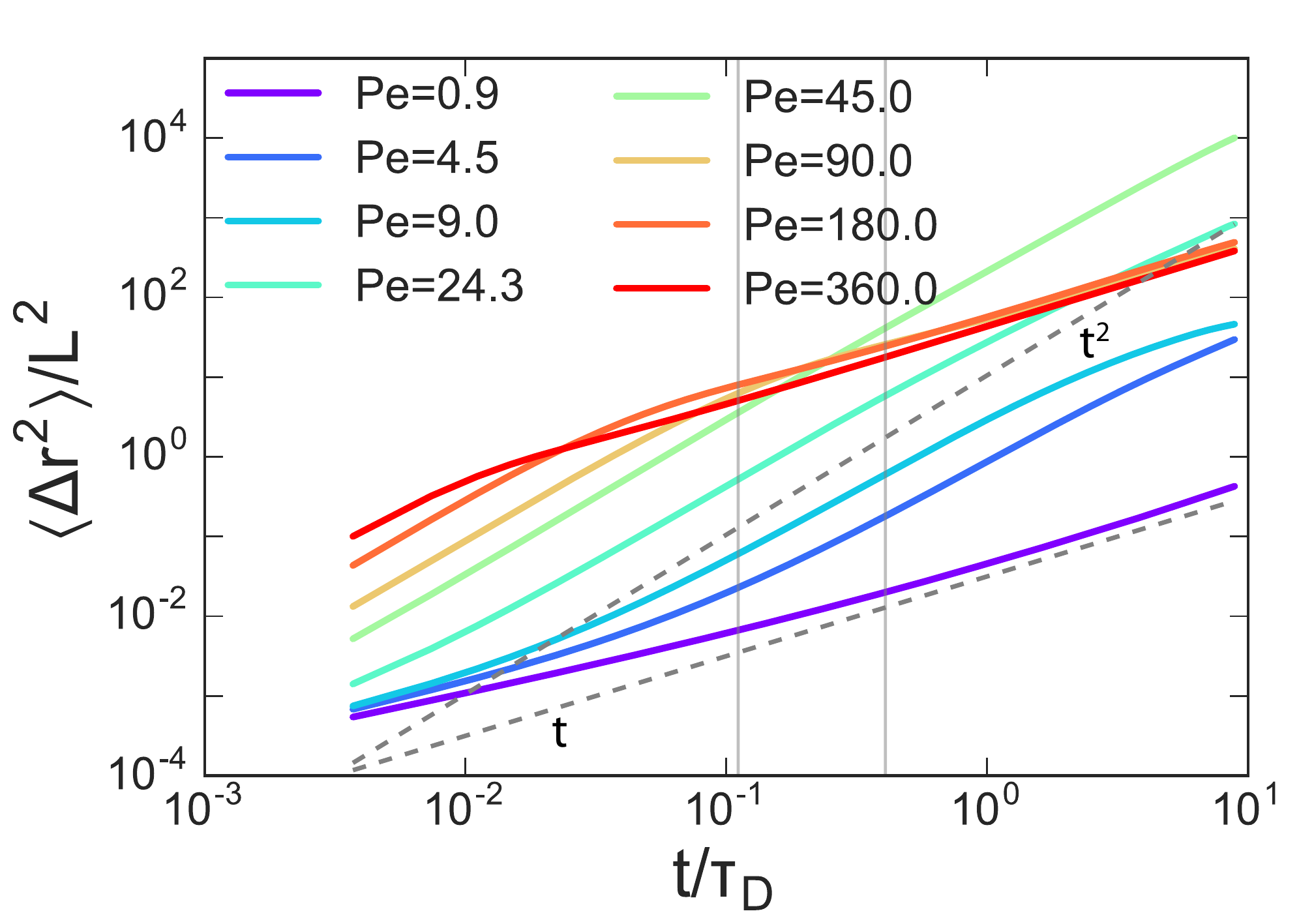}
 \caption{Mean-squared displacement of the center of mass of filaments as a function of lag time at $\phi = 0.8$, $a = 30$, $\xi_p/L = 0.2$ for various $Pe$. The gray-dashed lines indicate slopes of $1$ and $2$, depicting diffusive and ballistic motion, respectively. The gray vertical lines denote the range considered in the calculation of the slope in Fig. \ref{fgr:high_dens_phase_diagram}.}
 \label{fgr:high_dens_msd}
\end{figure}

To quantify the dynamics towards turbulence at high densities, we calculate the mean squared displacement (see Fig. \ref{fgr:high_dens_msd} for MSD of low persistence length filaments for various activities). The small displacements of stacks of filaments against one another in the jamming phase results in a short-time subdiffusive regime that becomes diffusive at late times in the jamming phase ($Pe = 0.9$). With increasing activity, the dynamics become superdiffusive in the laning regime as the filaments retain their orientational memory for extended durations ($Pe = 9.0-24.3$). At higher levels of activity, in the active turbulence regime, filament motion becomes diffusive again at intermediate times ($Pe = 90.0$). 

\begin{figure}[h]
\centering
 \includegraphics[width=0.95\columnwidth]{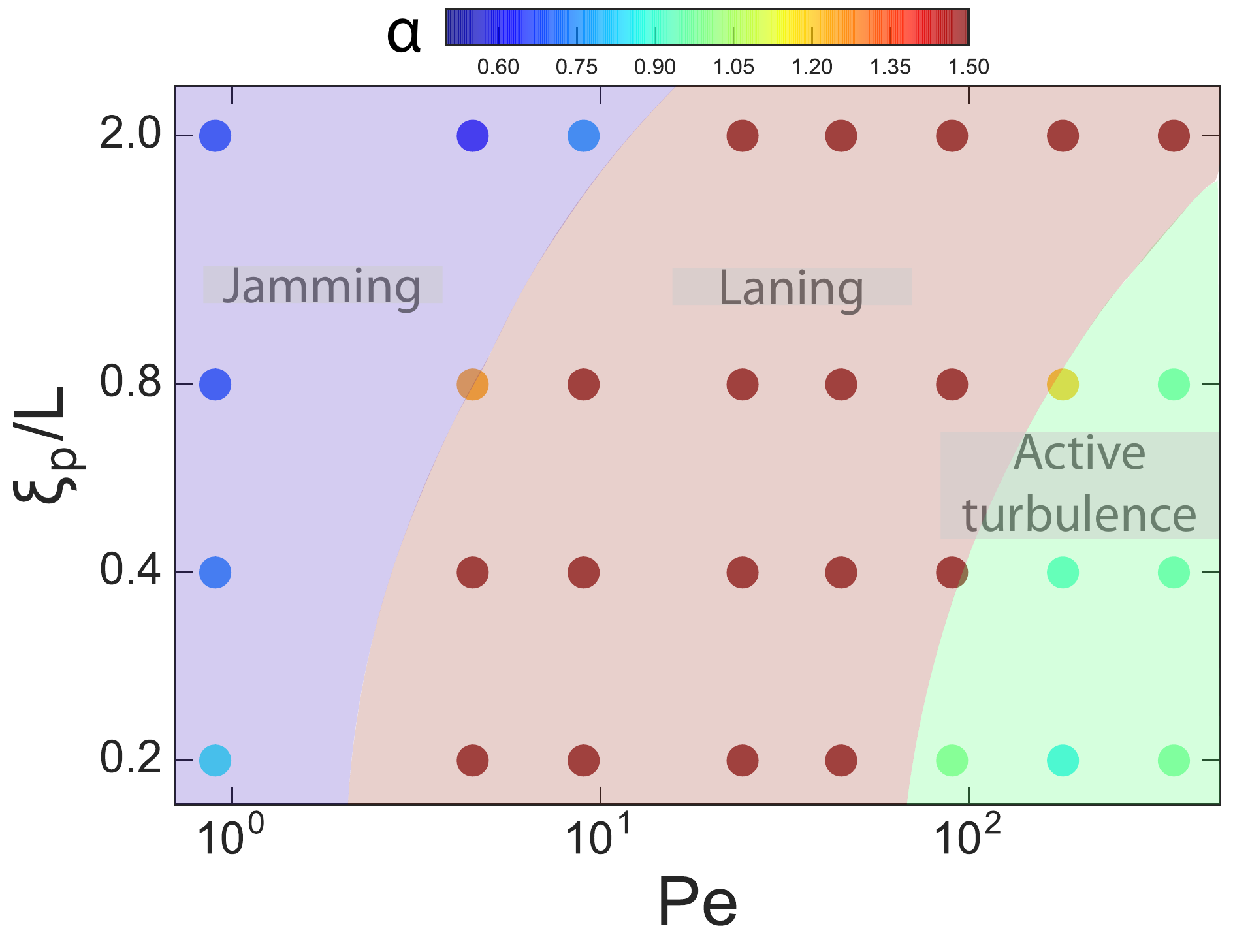}
 \caption{The phase diagram for $\phi = 0.8$ and $a = 30$ in terms of $\xi_p/L$ and $Pe$. The points denote individual simulation instances with the color code showing the exponent of MSD at intermediate times as depicted in Fig. \ref{fgr:high_dens_msd}.}
 \label{fgr:high_dens_phase_diagram}
\end{figure}

We use the intermediate-time (as defined by the gray vertical lines in Fig. \ref{fgr:high_dens_msd}) exponent of the MSD to construct a phase diagram (shown in Fig. \ref{fgr:high_dens_phase_diagram}). The active turbulence regime occurs at low persistence lengths and high levels of activity (in other words, at high flexure numbers), both of which promote lower values of effective persistence lengths and thereby lower degrees of correlated motion. Increasing effective persistence lengths (or decreasing flexure number), on the other hand, promotes jamming. 

\begin{figure}[h]
\centering
 \includegraphics[width=0.95\columnwidth]{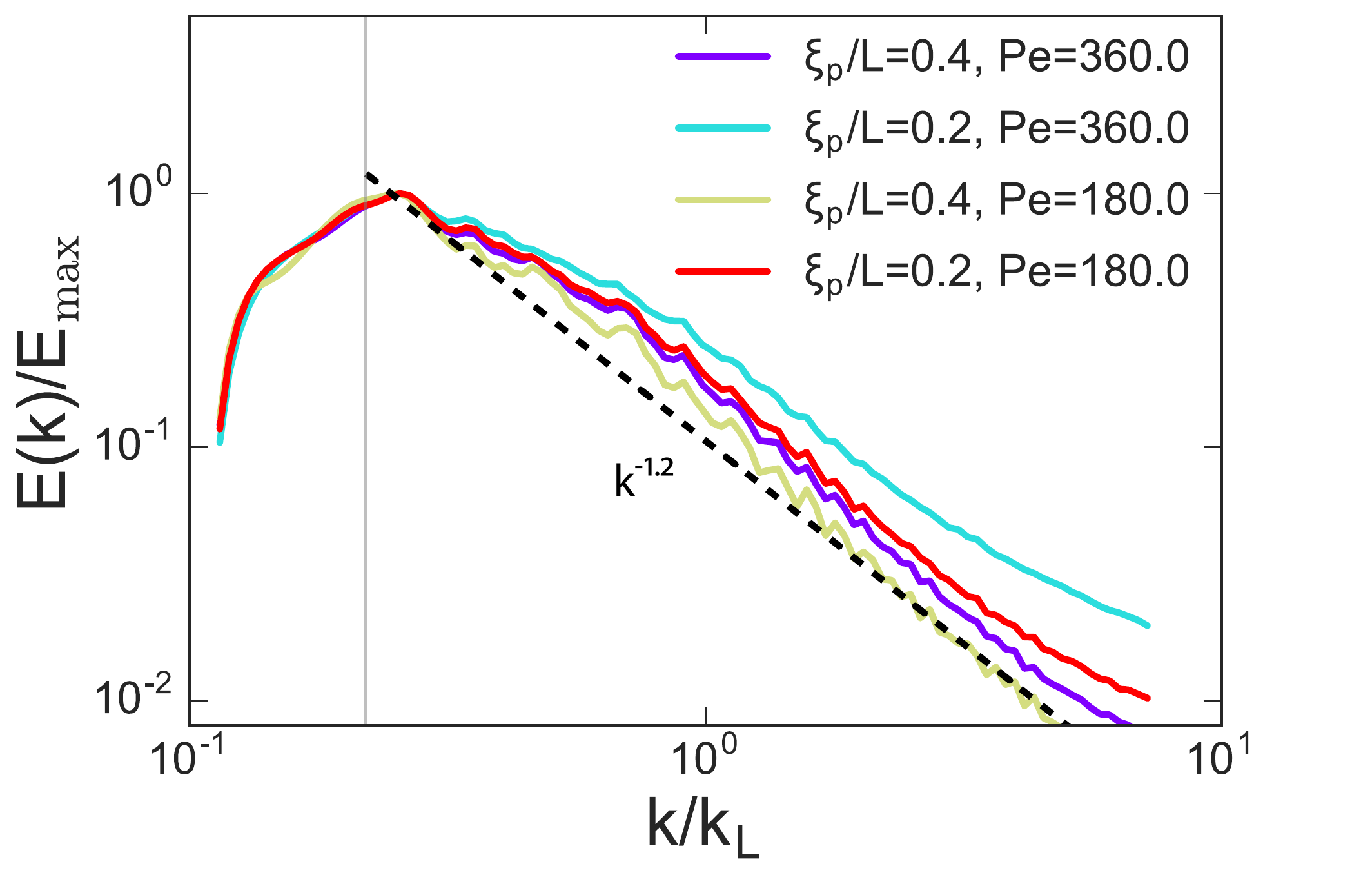}
 \caption{Kinetic energy spectrum $E(k)$, normalized by its maximum value $E_\mathrm{max}$ as a function of wavenumbers, normalized by the wavenumber corresponding to the filament length, $k_L = 2\pi/L$. The fit is a power law with exponent $-1.2 \pm 1$. The gray vertical line indicates half the simulation box length, corresponding to the largest length scale of the system.}
 \label{fgr:energy_spectrum}
\end{figure}

To characterize the active turbulence phase further, we calculate the kinetic energy spectrum, which can be obtained from the Fourier transform of spatial velocity correlations as
\begin{equation}
E(k) = \frac{k}{2\pi}\int\mathrm{d}^2R\mathrm{e}^{-i\mathbf{k}\cdot\mathbf{R}}\langle \mathbf{v}(t,\mathbf{r})\cdot\mathbf{v}(t,\mathbf{r}+\mathbf{r+R})\rangle ,
\label{eqn:kinetic_energy_spectrum}
\end{equation}
in two dimensions.\cite{batchelor1953theory, wensink2012meso} In the
active-turbulence regime, the kinetic energy accumulates in the
largest length scales and decreases toward lower length scales with a
power law $k^{-\alpha}$ with exponent $\alpha \approx 1.2 \pm 0.1$, see
Fig. \ref{fgr:energy_spectrum}. 
When normalized by the maximum kinetic energy, all curves roughly fall on top of each other, indicating that changing activity or rigidity does not lead to a significant change in the spectral behavior. Instead, it causes a change in the velocity magnitude. This is somewhat similar to the previously reported collapse of the spatial velocity correlations in the turbulent state of active nematics, when normalized by the mean-squared velocity.\cite{thampi2013velocity, giomi2015geometry}

\section{Discussion}

The extended and flexible nature of filamentous objects allows
deformations and self-interactions. This additional degree of freedom
enhances the importance of the behavior of individual filaments in the
collective dynamics. When the individual filaments have low rigidity,
filaments are bent, inhibiting their extension and directionality of motion. The resulting ensemble is weakly interacting,  as embodied in the gas-like dynamics of spirals and gas of clusters regimes. Stiff filaments, on the other hand, have extended rod-like shapes. They form larger and more persistent clusters that are strongly cooperative in dynamics. Filaments inside giant clusters not only point in the same direction, but also follow the change in orientation of other constituent filaments. In this way, spirals and giant clusters are results of the interplay between the extended and flexible nature of the filaments. 

The change from the spiralling state, where filament motion is constrained, to the clustering state, where filaments move in groups in an aligned fashion, represents dramatically different routes as to how propulsion force induces structure formation. In the clustering state, the propulsion force driving the system out of equilibrium is used for the movement of filaments whereas,  in the spiralling case, it is used for shape change to sustain a slowed-down steady state. It is indeed peculiar for an active system to turn inactive by a shape change without the involvement of an external cue. Actin filaments and microtubules in motility assays are found to exhibit this type of frozen steady state notwithstanding their activity.\cite{Schaller_Frozen_2011}

Activity enhances the ability of filaments to explore the rich configurational space provided by flexibility and aspect ratio. Therefore, the resultant phase space displays a spectrum of phases that mimic the behavior of other non-equilibrium systems. In the gas-of-spirals phase, filaments effectively perform thermal motion analogous to an ensemble of passive point particles. In the gas-of-clusters regime, on the other hand, filaments organize in small and short-lived clusters, reminiscent of the dynamical clusters observed for self-propelled point particles.\cite{Buttinoni_Dynamical_2013, Abkenar_Collective_2013, Wysocki_Cooperative_2014} Stiff filaments behave like self-propelled rigid rods in forming structures like giant clusters. In this respect, the collective dynamics of active semiflexible filaments provides a rich framework where different non-equilibrium behavior can be accessed. This can prove useful both in experimental applications and in distinguishing the rich behavior of biopolymers and filamentous structures in biology. 

The changing cluster dynamics with activity and flexibility can be
understood by analysing the collisions (see Movie M1 in the ESI).
 An increase in activity leads to an increased number of collisions which trap filaments for a duration that is proportional with their persistence in orientation. As a result, filaments form clusters with increasing activity due to collision-mediated self-trapping. However, the effect of collisions depends strongly on the flexibility of filaments. When filaments are stiff, they are structurally more elongated. In this regime, colliding filaments align in parallel or anti-parallel directions depending on the angle of collision being acute or obtuse, respectively. Flexible filaments, on the other hand, are easier to be bent upon collisions, which makes them reorient frequently. Therefore, clustering gets enhanced with increasing rigidity, but decreases again at high levels of activity. It can be argued that the cluster formation is a density-driven effect for flexible filaments, hence the formation of small and transient clusters, whereas it is an alignment-driven effect for stiff filaments, hence the large and persistent clusters.  

It is important to highlight the differences between self-propelled
rigid rods and semiflexible filaments. Stiff filaments are elongated
in a way that resembles rigid rods. Hence, both systems form giant
clusters as a result of alignment upon collisions. However,
flexibility of filaments still reflects itself even for
$\xi_p>>L$. Filaments can penetrate through antagonistically oriented
clusters, as they can bend, which alters the stress accumulation
mechanism in the ensemble. Therefore, some of the phases observed for
active rigid rods, like giant jammed structures,\cite{Yang_Swarm_2010,Weitz_Self_2015}
disappear for semiflexible filaments. 
A similar behavior of non-monotonic cluster size with self-propulsion has also been observed in 
ensembles of self-propelled overlapping rods and self-propelled colloids with short-range 
attractive interactions;\cite{Mani_Effect_2015, Mognetti_Living_2013, Abkenar_Collective_2013}
however, the underlying mechanisms are different.

Another important difference of semiflexible filaments and rigid rods is
that, when the leading tip of a filament turns by bending, the body of
the filament bends in the same direction. As a result of this ease in
reorientation, stiffer filaments can easily rotate even at very high
rigidities. Consequently, giant clusters of filaments are often in
curved and meandering form and change their direction of motion
frequently (see Movie M4 in the ESI). Such clustering dynamics is very similar 
to the dynamics of bird flocks and fish schools, where the collective dynamics 
is set by a number of leaders.\cite{Nagy_Hierarchical_2010, Cavagna_Scale_2010} 

The observed phases in the active semiflexible filament collective can be used in a tunable fashion. 
The system can be switched from a moving state with aligned filaments moving together to a frozen 
steady state with coiled filaments. This parameter-dependent effect can be used as a tunable switch in micro- and nano-technology, for which the first steps have already been made.\cite{Heuvel_Motor_2007} Besides the possible technological function, this type of self-organization also has a functional role in living matter. In a plant cell, microtubules form cortical arrays, which are curved and rotating structures in a two-dimensional plane along the cell wall, to provide stability and growth to the cell wall.\cite{Chan_Cortical_2007} Slender bacteria are observed to form spiral structures through mechanical interactions when put in a bath of shorter cells.\cite{Lin_Dynamics_2013} With their structural and dynamical resemblance to active semiflexible polymers, there is a plethora of biological systems like actin filaments and microtubules on molecular motor carpets, which can self-organize into spirals or into rotating clusters based on their level of rigidity, activity and aspect ratio. Our work can help in construction of such \emph{in-vitro} experiments. Additional effects like hydrodynamic interactions may need to be included in the interpretation of experimental results on microtubules in motility assays.\cite{DeCamp_Nat_2015} Simulation studies of active polar semiflexible filaments in the low-density regime indicate a 
qualitatively similar behavior with and without hydrodynamics.\cite{Jiang_Hydrodynamic_2014, Jiang_Motion_2014} 

\section{Summary}

We have studied the collective dynamics of active semiflexible filaments in a two-dimensional system with steric interactions. With a minimal active polymer model, we are able to capture rich dynamical behavior. The collective dynamics has the hallmarks of a passive homogeneous melt for low propulsion strengths. When the propulsion is increased, filaments organize in small and transient clusters similar to the dynamical clusters observed in active point particles. As the rigidity is increased, we find that filaments form giant clusters like self-propelled rigid rods, when they are propelled at intermediate levels of activity. Compared with rods, clustering is not homogeneous with activity due to the deformation of filaments upon collisions. The aspect ratio of filaments plays an important role in the dynamics of strongly propelled and flexible filaments. Such filaments self-organize in different configurations of spiral aggregates in which they perform diffusive motion. At high densities, we distinguish three phases: At low activity and low rigidity, filaments are jammed in their initial configurations. With increasing activity, filaments break out of the cages. At intermediate levels of activity, filaments form nematic lanes. At high activity, laning becomes instable and filaments move in smaller nematic bands. The collisions between these nematic bends result in a turbulent regime.

\section*{Acknowledgments}

The authors gratefully acknowledge helpful discussions with Thorsten Auth and Arvind Ravichandran, financial support by the Deutsche Forschungsgemeinschaft through the priority program SPP 1726 on ''Microswimmers'', and a computing-time grant on the supercomputer JURECA at J\"ulich Supercomputing Centre (JSC).

\balance

\bibliography{refs.bib} 
\bibliographystyle{rsc} 

\end{document}